\documentclass[11pt]{article}

\usepackage{amssymb}

\newcommand{\bea}{\begin{eqnarray}}
\newcommand{\eea}{\end{eqnarray}}
\newcommand{\beq}{\begin{equation}}
\newcommand{\eeq}{\end{equation}}

\newcommand{\cL}{{\cal L}}

\def\l{\left}
\def\r{\right}

\begin{document}

\begin{center}{\tt
\hfill LPT Orsay, 05-53 \\
\hfill ROMA-1412/05} \\
\begin{center}

\vspace{3.2cm}

{\LARGE\bf The Ademollo--Gatto theorem 

\vspace{0.2cm}
 for lattice semileptonic decays}

\end{center}

\vspace{1.cm}

{\large \sc D. Be\'cirevi\'c$^{a}$, G. Martinelli$^{b}$ and G. Villadoro$^{b}$} \\

\vspace{0.8cm}

${}^a\!\!$
{\sl Laboratoire de Physique Th\'eorique, Universit\'e Paris Sud, \\
           Centre d'Orsay, F-91405 Orsay-Cedex, France.} \\
\vspace{.3cm}
${}^b\!\!$
{\sl Dipartimento di Fisica, Universit\`a di Roma ``La Sapienza'', } \\ 
 {\sl          and INFN, Sezione di Roma, P.le A. Moro 2, I-00185 Rome, Italy.} \\
\vspace{.3cm}

\end{center}

\vspace{0.8cm}

\centerline{\bf Abstract}
\begin{quote}\small

We present the results of the calculation  of the $K_{\ell 3}$ semileptonic form factor at zero momentum transfer, 
$f_+(0)$,  obtained at one-loop in  partially quenched Chiral Perturbation Theory     
(with either $n_f=2$, or $n_f=3$,  and with generic valence and sea quark masses).  
We show that for $n_f=2$,  when the masses of the valence and sea light quarks are equal, 
the correction is of ${\cal O}[(M_K^2-M_\pi^2)^3]$.  The formulae presented here can be useful 
for the mass extrapolation of the results obtained in lattice simulations to the physical point. 

\end{quote}

\newpage

\baselineskip 5mm

\section{Introduction}
\label{sec:intro}
In the last two years  we assisted to a renewed interest in theoretical calculations of the semileptonic 
form factor $f_+(q^2)$ relevant to the extraction of $\vert V_{us}\vert$ from $K \to \pi \ell \bar \nu_\ell$  
($K_{\ell3}$) decays~\cite{bij}-\cite{altrikl3}.
In particular it  has been shown that in lattice simulations  the form factor at zero recoil,
$f_+(q^2=0)$,   can be extracted with  the percent precision that is required  for making a meaningful 
determination  of  $\vert V_{us}\vert$~\cite{noikl3}. Although many systematic uncertainties must still  
be reduced,  by  performing unquenched calculations  at lower quark masses and on several lattice spacings, 
the calculation  of ref.~\cite{noikl3}  triggered a new wave of activity and the quality of the  results is rapidly 
improving~\cite{altrikl3}. 

A key  observation which allows to reach a good theoretical control of these transitions is the Ademollo--Gatto 
theorem~\cite{ag}, which states that the $K_{\ell3}$ form factors  $f_+(q^2)$ and $f_0(q^2)$ at zero momentum 
transfer  ($q^2=0$) are renormalised only by terms of second order in the breaking of the  $SU(3)$  flavour symmetry.  
Besides, chiral perturbation theory (ChPT)  provides an excellent tool to analyse the dependence of $f_{+,0}(0)$ 
 on the meson (quark) masses,  and a guidance for the extrapolation of the lattice form factors to 
the physical point.  Following Leutwyler and Roos it is convenient to express the form factor in the form~\cite{LeuRo} 
\beq 
f_+(0) = 1 + f_2 + f_4 + \dots \, , \label{eq:LR} 
\eeq
where $f_n = {\cal O}[M_{K,\pi}^n/(4\pi f_\pi)^n]$ are the terms arising at higher  
orders in ChPT. Because of the Ademollo--Gatto theorem, the first non-trivial term in the chiral expansion, 
$f_2$,  does not receive contributions from local operators  
appearing in the effective theory and can be computed unambiguously in terms of $M_K$, $M_\pi$ and $f_\pi$~\cite{LeuRo}.

Lattice calculations of the $K_{\ell3}$ form factors have been done in quenched and partially quenched 
($n_f=2$) QCD. In the latter case simulations are performed with ``valence'' quark masses  equal to or different from ``sea'' quark masses.  In such a situation a number of subtleties related to the validity of the Ademollo--Gatto 
theorem arise. In this paper we discuss the applicability of Ademollo--Gatto theorem in various situations 
(quenched, partially quenched and fully unquenched), and give the main expressions for $f_2$ in each case.
These formulae are important for the extrapolation of the form factors to the physical point. 
In the following we will always work in the isospin symmetric limit, with the mass of the strange quark 
($m_s$) different from the mass of the light quarks ($m_d=m_u$).
\section{Quenched and unquenched formulae}
\label{sec:qeuq}
In this section we give a brief summary of the known  results for $f_2$, namely in full QCD and its quenched approximation.
\begin{itemize} 
\item[$\circ$] {\bf Full QCD} \\
In the isospin-symmetric limit, within full QCD, 
the expression of the leading chiral correction $f_2$ is~\cite{LeuRo}
\begin{equation}
f_2={3\over 2} H_{\pi K} + {3 \over 2} H_{\eta K} \; ,
\label{eq:f2-full}
\end{equation}
where 
\begin{equation}
H_{PQ}= -{1 \over 64 \pi^2 f_\pi^2} \left[ M_P^2+M_Q^2+ {2 M_P^2 M_Q^2
    \over M_P^2 - M_Q^2} \ln {M_Q^2 \over M_P^2} \right] \; \; .
\label{eq:HH}
\end{equation}
Note that $f_2$ is completely specified in terms of pseudoscalar
meson masses and decay constants ($f_\pi \approx 132$ MeV), it is negative
($f_2 \approx -0.023$ for physical masses), as implied by unitarity \cite{LeuRo,FLRS}, and  vanishes 
as $(M_K^2-M_\pi^2)^2/(f_\pi^2 M_K^2)$ in the SU(3) limit,  following the combined constraints of 
chiral symmetry and the Ademollo--Gatto theorem. 

\item[$\circ$] {\bf Quenched  QCD} \\
The structure of chiral logarithms appearing in eqs.~(\ref{eq:f2-full})--(\ref{eq:HH}),
is valid only in the full theory. In the quenched theory, instead, the leading (unphysical)  logarithms are 
 those entering the one-loop functional of qChPT~\cite{cp-q,bg-q,Sharpe}. 
$f_2$ in the quenched case was first computed in ref.~\cite{noikl3}.\par 
Normalising the lowest-order qChPT Lagrangian as in ref.~\cite{cp-q}, with a quadratic term 
for the singlet field $\Phi_0={\rm str}(\Phi)$ chosen as
\beq
\biggl. \cL^q_2 \biggr\vert_{\Phi^2_0} = \frac{\alpha}{6} D_\mu \Phi_0 D^\mu \Phi_0 
- \frac{M_0^2}{6} \Phi_0^2~,
\eeq 
the result is 
\beq
f_2^q  = H^q_{\pi K} +  H^q_{(s\bar{s}) K}~,
\label{eq:f2-quenched}
\eeq
where
\beq
 H^q_{P K} =  \frac{M_K^2}{96 \pi^2 f_\pi^2}\l [ 
  \frac{ M_0^2 ( M_K^2+M_{P}^2) - 2 \alpha M_K^2 M_{P}^2  }
  { {\left( M_K^2-M_{P}^2 \right) }^2}\,\log \l (\frac{M_{K}^2}{M_{P}^2}\r) -\,\alpha \r ]\,,
\eeq
with $M_{(s\bar{s})}^2=2M_K^2-M_\pi^2$. As anticipated, the one-loop result in 
eq.~(\ref{eq:f2-quenched}) is finite because of the
Ademollo--Gatto theorem, which is still valid in the quenched 
approximation~\cite{cp-q}, and thus the absence of contributions from local
operators in $f^q_2$.  A proof  that the Ademollo--Gatto theorem 
(and more generally the Sirlin's relation \cite{s-sr}) holds within qChPT beyond the one-loop level  
can easily be obtained by applying the functional formalism to the demonstration  
in ref.~\cite{s-sr}. The latter needs only flavour symmetries  for valence quarks
which hold on the lattice also in the quenched case.

It is worth emphasising that the nature of the SU(3) 
breaking corrections in the quenched theory is completely different 
from that of full QCD: only contributions coming from the mixing 
with the flavour singlet state are present and one finds $f^q_2>0$, which is a 
signal of the non-unitarity of the theory. 
For typical values of the singlet parameters ($M_0 \approx 0.6$ GeV
and $\alpha \approx 0$~\cite{bardeen}) and for the physical values of pion and kaon masses,
one finds $f^q_2 \approx + 0.022$. 
\end{itemize}

\section{Partially Quenched results}
\label{sec:pq}

In this section we give the new results for various set-ups relevant to partially quenched QCD. 
We have used the partially quenched  ChPT Lagrangian defined in refs.~\cite{bg-pq,ss}, and 
work with two sea quark masses ($m_s^{(S)}$,   $m_d^{(S)}$), and two valence ones ($m_s^{(V)}$, $m_d^{(V)}$).
We stress again that we always work in the exact isospin limit, i.e., $m_u=m_d$.
The meson masses, at leading order in ChPT,  read 
\bea 
M^2_\pi &= & 2 B_0 \, m_d^{(V)} \, , \quad  \quad  M^2_K =   B_0 \, \left(m_s^{(V)}  + m_d^{(V)}  \right) \, ,  \nonumber \\
M^2_{dd} &= & 2 B_0 \, m_d^{(S)} \, , \quad  \quad  M^2_{ss} =   2 B_0 \, m_s^{(S)}  \, , \label{eq:mesmass}  
\eea
where $B_0$ is the chiral condensate (more precisely, $B_0=-2\langle \bar q q\rangle / f_\pi^2$).
In the appendix we give the complete formula for $f_2^{pq}$, as obtained with $3$ 
dynamical flavours and four quark masses enumerated above. Here we focus onto the 
limits that are particularly interesting to the situations encountered in the partially quenched 
QCD simulations on the lattice.

Like in the cases of full and quenched QCD, also in the partially quenched theory the Ademollo--Gatto
theorem holds non-perturbatively to all orders in the chiral expansion. However, 
the generic structure of the lowest order correction in ChPT, expanded in the  mass difference 
of the valence quark masses,  reads
\beq  
f^{pq}_2    = \left[ \frac{g_1}{m_s^{(S)}} + g_2 \, \left( m_d^{(S)}-m_d^{(V)}\right) \right]  \times
\left( m_s^{(V)}-m_d^{(V)} \right)^2 + {\cal O}\left[( m_s^{(V)}-m_d^{(V)})^3\right] \, , 
\eeq 
where $g_1$ and $g_2$ are functions of the valence and sea   quark (meson) masses.   
Thus we find that in the 
 partially quenched theory with $n_f=2$, which is obtained by sending $ m_s^{(S)} \to \infty$, the 
correction is at least of the third order in $m_s^{(V)}-m_d^{(V)}$  if the valence and sea 
light quark masses  are the same. This  is only an accident however: at the next order 
in the chiral expansion, the  corrections of  $ {\cal O} \left[( m_s^{(V)}-m_d^{(V)})^2\right]$, 
due to the  effect of  the higher-dimensional  local operators,  will appear.   This implies that 
a numerical analysis of the mass dependence of the form factor $f_{+,0}(0)$ in the $n_f=2$ case 
and with $m_d^{(S)}=m_d^{(V)}$,  could determine the constants quite precisely since the 
leading non-analytic corrections from $f_2^{pq}$ are suppressed by this enhanced AG effect.

\noindent In the following we give the resulting expressions in two important cases:
\begin{itemize}
\item[$\circledcirc$]  {\bf  \boldmath{$n_f=2$} non-degenerate valence and sea light quarks} \\
In this case we have 
\bea  f^{pq}_2 &=& - \frac{2\,M_K^2 + M_{dd}^2}{32\,{\pi }^2\,f_\pi^2} + 
  \frac{M_K^2\,\left[ M_\pi^2\,M_{dd}^2 + 
       M_K^2\,\left(M_{dd}^2 -2\,M_\pi^2 \right)  \right] \,
     \log \l(\frac{M_K^2}{M_\pi^2}\r)}{64\,{\pi }^2\,f_\pi^2\,{\left( M_K^2 - M_\pi^2 \right) }^2} +  \nonumber \\ && +
  \frac{M_K^2\,\left[ \l ( 2\,M_K^2-M_{dd}^2 \r)\,\left( 2\,M_K^2 - M_\pi^2 \right)  - 
       M_K^2\,M_{dd}^2 \right] \,
     \log \l(2 - \frac{M_\pi^2}{M_K^2}\r)}{64\,{\pi }^2\,f_\pi^2\,{\left( M_K^2 - M_\pi^2 \right) }^2}+  \nonumber \\ && 
   + \frac{\left( 2\,M_K^2 - M_\pi^2 + M_{dd}^2 \right) \,
     \left( M_\pi^2 + M_{dd}^2 \right) \,
     \log \l(\frac{2\,M_K^2 - M_\pi^2 + M_{dd}^2}{M_\pi^2 + M_{dd}^2}\r)}{64\,{\pi }^2\,f_\pi^2\,
     \left( M_K^2 - M_\pi^2 \right) } \, , 
\eea 
which in the small $M_K^2 -M_\pi^2$ limit becomes 
\beq f^{pq}_2 = - \frac{{\left( M_K^2 - M_\pi^2 \right) }^2\,\left( M_K^2 - M_{dd}^2 \right) \,
     \left( 3\,M_K^2 + M_{dd}^2 \right) }{192\,{\pi }^2\,f_\pi^2\,M_K^4\,
     \left( M_K^2 + M_{dd}^2 \right) }+ {\cal O}\left[(M_K^2 - M_\pi^2)^4\right] \, .\nonumber  
\eeq
Note that only even powers of $M_K^2 - M_\pi^2 $ appear in the expansion, as in the quenched case. 
The odd powers are hidden in the factor $M_K^2-M_{dd}^2$.

\item[$\circledcirc$]  {\bf  \boldmath{$n_f=2$} degenerate valence and sea light quarks} \\
By taking $M_{dd} = M_\pi$ we get the case with $ m_d^{(S)}=m_d^{(V)}$. The correction is now cubic 
in the $SU(3)$ breaking, namely, 
\bea f^{pq}_2 &=&  - \frac{2\,M_K^2 + M_\pi^2}{32\,{\pi }^2\,f_\pi^2} - 
  \frac{3\,M_K^2\,M_\pi^2\,\log \l(\frac{M_\pi^2}{M_K^2}\r)}
   {64\,{\pi }^2\,f_\pi^2\,\left( M_K^2 - M_\pi^2 \right) } + \nonumber\\
&& +  \frac{M_K^2\,\left( 4\,M_K^2 - M_\pi^2 \right) \,
     \log \l(2 - \frac{M_\pi^2}{M_K^2}\r)}{64\,{\pi }^2\,f_\pi^2\,\left( M_K^2 - M_\pi^2 \right) } \, , \label{eq:nf2d}
\eea
 which after expanding in $M_K^2-M_\pi^2$ leads to
\bea f^{pq}_2 =  - \frac{{\left( M_K^2 - M_\pi^2 \right) }^3}{96\,{\pi }^2\,f_\pi^2\,M_K^4} + 
  {\cal O}\left[{(M_K^2 - M_\pi^2)}^4 \right]\, , \eea 
thus showing the suppression of the SU(3)-breaking corrections at this order.
In particular, in this case, the AG quadratic correction extracted from lattice simulation 
of the $K\to\pi$ vector form factor, will be free from the $f_2$ contributions and will start with $f_4$ 
where analytic contributions are present.
\end{itemize}

\section{Conclusion} 
In this paper we discussed the  leading chiral corrections to the  $K_{\ell 3}$ form factor $f(0)$, that are 
protected by the Ademollo--Gatto theorem and thus unambiguously computable in all three forms 
of ChPT, i.e. the ones corresponding to the full, partially quenched and quenched QCD. 
We provide the formulae for the partially quenched case which are needed for the mass 
extrapolation of currently accessible $f(0)$ computed on the lattice to the physical  kaon and pion masses.  The complete formula, with generic sea and valence quark (meson) masses, is given in Appendix, wheres the  case  of $n_f=2$ is discussed in more detail in the text.  We show that in 
the latter case with the valence and sea light quarks being degenerate in mass,  the form factor $f(0)$ 
is free from $f_2^{pq}$ correction, thus allowing for ever cleaner determination of $f_4$ from the 
lattice.

\section*{Acknowledgment} G.M. and G.V. warmly thank the Benasque Center for Science 
for hospitality where, during the workshop ``Matching light quarks to hadrons - 2004", this work 
has been initiated.  D.B. thanks the CERN (theory division) for hospitality, where this work has 
been finalised.

\section*{Appendix} 
In this appendix we give the formula for $f^{pq}_2$ for $n_f=3$ and generic valence and sea quark masses. 
The case $n_f=2$  discussed in the text is readily obtained by sending $m_s^{(S)} \to \infty$.  The full 
case is recovered by taking $m_s^{(S)}=m_s^{(V)}$ and $m_d^{(S)}=m_d^{(V)}$, 
corresponding to $M_{ss}^2= 2\, M_K^2-M_\pi^2$ and $M^2_{dd}= M_\pi^2$.  
{\footnotesize
\bea f^{pq}_2&=&  M_K^2\,\Bigl[ \left( 2\,M_K^2 - M_\pi^2 \right) \,
      \Bigl ( 6\,M_K^2\,{\left( 2\,M_K^2 - M_\pi^2 \right) }^2  
        \nonumber \\ && 
	   - M_{ss}^2\,\left( \left( 2\,M_K^2 - M_\pi^2 \right) \,
            \left( 11\,M_K^2 - M_\pi^2 \right)  + 4\,M_K^2\,M_{ss}^2 \right)  \Bigr)  \nonumber \\ && 
    - 2\,\Bigl ( \left( 5\,M_K^2 - M_\pi^2 \right) \,{\left( 2\,M_K^2 - M_\pi^2 \right) }^2 - 
        3\,\left( 2\,M_K^2 - M_\pi^2 \right) \,\left( 3\,M_K^2 - M_\pi^2 \right) \,
         M_{ss}^2 \nonumber  \\ && 
		 + \left( 3\,M_K^2 - M_\pi^2 \right) \,M_{ss}^4 \Bigr ) \,M_{dd}^2 + 
      \Bigl ( M_\pi^2\,M_{ss}^2 + M_K^2\,
         \left( 4\,M_K^2 - 2\,M_\pi^2 - 3\,M_{ss}^2 \right)  \Bigr) \,M_{dd}^4 \Bigr] \times  \nonumber \\ && 
		 \times \frac{\log \left(\frac{2\, M_K^2 - M_\pi^2}{M_\pi^2}\right)}
   {32\,{\pi }^2\,f_\pi^2\,{\left( M_K^2 - M_\pi^2 \right) }^2\,
     {\left( 3\,M_\pi^2 + 2\,M_{ss}^2 + M_{dd}^2 -6\,M_K^2 \right) }^2}  -  \nonumber \\ &&
   - \frac{M_K^4\,\left( M_K^2 - M_{ss}^2 \right) \,
     \left( M_K^2 - M_{dd}^2 \right) \,\log \left(\frac{M_K^2}{M_\pi^2}\right)}{8\,{\pi }^2\,f_\pi^2\,
     {\left( M_K^2 - M_\pi^2 \right) }^2\,
     \left( 3\,M_K^2 - 2\,M_{ss}^2 - M_{dd}^2 \right) }   + \nonumber  \\ && 
 + \frac{\left( 2\,M_K^2 - M_\pi^2 + M_{ss}^2 \right) \,
     \left( M_\pi^2 + M_{ss}^2 \right) \,
     \log \left(\frac{2\,M_K^2 - M_\pi^2 + M_{ss}^2}{M_\pi^2 + M_{ss}^2}\right)}{128\,{\pi }^2\,f_\pi^2\,
     \left( M_K^2 - M_\pi^2 \right) }  +  \nonumber \\ &&+
  \frac{\left( 2\,M_K^2 - M_\pi^2 + M_{dd}^2 \right) \,
     \left( M_\pi^2 + M_{dd}^2 \right) \,
     \log \left(\frac{2\,M_K^2 - M_\pi^2 + M_{dd}^2}{M_\pi^2 + M_{dd}^2}\right)}{64\,{\pi }^2\,f_\pi^2\,
     \left( M_K^2 - M_\pi^2 \right) } -\nonumber \\ &&-
  \frac{3\,M_K^2\,{\left( M_K^2 - M_\pi^2 \right) }^2\,
     {\left( M_{ss}^2 - M_{dd}^2 \right) }^2\,\left( 2\,M_{ss}^2 + M_{dd}^2 \right) \,
     \log \left(\frac{2\,M_{ss}^2 + M_{dd}^2}{3 M_\pi^2}\right)}{4\,{\pi }^2\,f_\pi^2\,
     \left( 3\,M_K^2 - 2\,M_{ss}^2 - M_{dd}^2 \right) \,
     {\left(  2\,M_{ss}^2 + M_{dd}^2 -3\,M_\pi^2 \right) }^2\,
     {\left( 3\,M_\pi^2 + 2\,M_{ss}^2 + M_{dd}^2 -6\,M_K^2  \right) }^2}    
     \nonumber \\ &&+
  \frac{26\,M_K^4-\,\left(  2\,M_{ss}^2 + M_{dd}^2 +3\,M_\pi^2\right) 
     \left( M_{ss}^2 + 2\,M_{dd}^2 \right)} {64\,{\pi }^2\,f_\pi^2\,
     \left( 3\,M_\pi^2 + 2\,M_{ss}^2 + M_{dd}^2 -6\,M_K^2\right) }- \nonumber \\ && -
     \frac{M_K^2\,\left( 39\,M_\pi^4 - 8\,M_{ss}^4 - 
        18\,M_\pi^2\,\left( M_{ss}^2 + 2\,M_{dd}^2 \right)  + 
        M_{dd}^2\,\left( 18\,M_{ss}^2 + 5\,M_{dd}^2 \right)  \right) }{64\,{\pi }^2\,f_\pi^2\,
     \left( 3\,M_\pi^2 - 2\,M_{ss}^2 - M_{dd}^2 \right) \,
     \left( 3\,M_\pi^2 + 2\,M_{ss}^2 + M_{dd}^2  -6\,M_K^2 \right) } \,.
  \nonumber   \eea }

\end{document}